\documentclass[12pt]{article}
\begin{document}
\newcommand{\beq}{\begin{equation}}
\newcommand{\eeq}{\end{equation}}
\newcommand{\beqa}{\begin{eqnarray}}
\newcommand{\eeqa}{\end{eqnarray}}
\newcommand{\beqar}{\begin{eqnarray*}}
\newcommand{\eeqar}{\end{eqnarray*}}
\newcommand{\al}{\alpha}
\newcommand{\be}{\beta}
\newcommand{\del}{\delta}
\newcommand{\D}{\Delta}
\newcommand{\eps}{\epsilon}
\newcommand{\ga}{\gamma}
\newcommand{\Ga}{\Gamma}
\newcommand{\ka}{\kappa}
\newcommand{\nn}{\nonumber}
\newcommand{\inn}{\!\cdot\!}
\newcommand{\h}{\eta}
\newcommand{\ii}{\iota}
\newcommand{\kk}{\varphi}
\newcommand\F{{}_3F_2}
\newcommand{\la}{\lambda}
\newcommand{\La}{\Lambda}
\newcommand{\na}{\prt}
\newcommand{\Om}{\Omega}
\newcommand{\om}{\omega}
\newcommand\dS{\dot{\cal S}}
\newcommand\dB{\dot{B}}
\newcommand\dG{\dot{G}}
\newcommand\ddG{\ddot{G}}
\newcommand\ddB{\ddot{B}}
\newcommand\ddP{\ddot{\phi}}
\newcommand\dP{\dot{\phi}}
\newcommand{\p}{\phi}
\newcommand{\sig}{\sigma}
\renewcommand{\t}{\theta}
\newcommand{\z}{\zeta}
\newcommand{\ssc}{\scriptscriptstyle}
\newcommand{\eg}{{\it e.g.,}\ }
\newcommand{\ie}{{\it i.e.,}\ }
\newcommand{\labell}[1]{\label{#1}} %{\label{#1}} %
\newcommand{\reef}[1]{(\ref{#1})}
\newcommand\prt{\partial}
\newcommand\veps{\varepsilon}
\newcommand{\pol}{\varepsilon}
\newcommand\vp{\varphi}
\newcommand\ls{\ell_s}
\newcommand\cF{{\cal F}}
\newcommand\cA{{\cal A}}
\newcommand\cS{{\cal S}}
\newcommand\cT{{\cal T}}
\newcommand\cV{{\cal V}}
\newcommand\cL{{\cal L}}
\newcommand\cM{{\cal M}}
\newcommand\cN{{\cal N}}
\newcommand\cG{{\cal G}}
\newcommand\cH{{\cal H}}
\newcommand\cI{{\cal I}}
\newcommand\cJ{{\cal J}}
\newcommand\cl{{\iota}}
\newcommand\cP{{\cal P}}
\newcommand\cQ{{\cal Q}}
\newcommand\cg{{\it g}}
\newcommand\cR{{\cal R}}
\newcommand\cB{{\cal B}}
\newcommand\cO{{\cal O}}
\newcommand\tcO{{\tilde {{\cal O}}}}
\newcommand\bg{\bar{g}}
\newcommand\bb{\bar{b}}
\newcommand\bH{\bar{H}}
\newcommand\bX{\bar{X}}
\newcommand\bK{\bar{K}}
\newcommand\bA{\bar{A}}
\newcommand\bZ{\bar{Z}}
\newcommand\bxi{\bar{\xi}}
\newcommand\bphi{\bar{\phi}}
\newcommand\bpsi{\bar{\psi}}
\newcommand\bprt{\bar{\prt}}
\newcommand\bet{\bar{\eta}}
\newcommand\btau{\bar{\tau}}
\newcommand\bnabla{\bar{\nabla}}
\newcommand\hF{\hat{F}}
\newcommand\hA{\hat{A}}
\newcommand\hT{\hat{T}}
\newcommand\htau{\hat{\tau}}
\newcommand\hD{\hat{D}}
\newcommand\hf{\hat{f}}
\newcommand\hg{\hat{g}}
\newcommand\hp{\hat{\phi}}
\newcommand\hi{\hat{i}}
\newcommand\ha{\hat{a}}
\newcommand\hb{\hat{b}}
\newcommand\hQ{\hat{Q}}
\newcommand\hP{\hat{\Phi}}
\newcommand\hS{\hat{S}}
\newcommand\hX{\hat{X}}
\newcommand\tL{\tilde{\cal L}}
\newcommand\hL{\hat{\cal L}}
\newcommand\tG{{\widetilde G}}
\newcommand\tg{{\widetilde g}}
\newcommand\tphi{{\widetilde \phi}}
\newcommand\tPhi{{\widetilde \Phi}}
\newcommand\td{{\tilde d}}
\newcommand\tk{{\tilde k}}
\newcommand\tf{{\tilde f}}
\newcommand\ta{{\tilde a}}
\newcommand\tb{{\tilde b}}
\newcommand\tc{{\tilde c}}
\newcommand\tR{{\tilde R}}
\newcommand\teta{{\tilde \eta}}
\newcommand\tF{{\widetilde F}}
\newcommand\tK{{\widetilde K}}
\newcommand\tE{{\widetilde E}}
\newcommand\tpsi{{\tilde \psi}}
\newcommand\tX{{\widetilde X}}
\newcommand\tD{{\widetilde D}}
\newcommand\tO{{\widetilde O}}
\newcommand\tS{{\tilde S}}
\newcommand\tB{{\widetilde B}}
\newcommand\tA{{\widetilde A}}
\newcommand\tT{{\widetilde T}}
\newcommand\tC{{\widetilde C}}
\newcommand\tV{{\widetilde V}}
\newcommand\thF{{\widetilde {\hat {F}}}}
\newcommand\Tr{{\rm Tr}}
\newcommand\tr{{\rm tr}}
\newcommand\STr{{\rm STr}}
\newcommand\hR{\hat{R}}
\newcommand\M[2]{M^{#1}{}_{#2}}

\newcommand\bS{\textbf{ S}}
\newcommand\bI{\textbf{ I}}
\newcommand\bJ{\textbf{ J}}

%\begin{document}
\begin{titlepage}
\begin{center}

\vskip 2 cm
{\LARGE \bf Background independence of   \\ \vskip 0.75  cm  effective actions at critical dimension  }\\
\vskip 1.25 cm
   Mohammad R. Garousi\footnote{garousi@um.ac.ir}

\vskip 1 cm
{{\it Department of Physics, Faculty of Science, Ferdowsi University of Mashhad\\}{\it P.O. Box 1436, Mashhad, Iran}\\}
\vskip .1 cm
 \end{center}

\begin{abstract}
Recently, by explicit calculations at orders $\alpha',\alpha'^2,\alpha'^3$, it has been observed   that the effective action of string theory at the critical dimension is independent of the background for the closed spacetime manifolds. In this paper we speculate  that for the open spacetime manifolds, the effective action   is even independent of the character of the boundary, \ie the boundary couplings for timelike and spacelike boundaries are the same.
We support this proposal by calculating the boundary couplings in the bosonic string theory  at order $\alpha'$ for the spacelike boundary and show that they are the same as the couplings for the timelike boundary that have been recently found.
\end{abstract}
\end{titlepage}

\section{Introduction}

The critical string theory is an extension of the Einstein theory of general relativity which is consistent with the rules of quantum mechanics. As in the Einstein theory, one expects the string theory at the critical dimension $D=1+d$ to be background independent. In the low energy effective action, by the background independence we mean the coefficients of the independent  gauge invariant couplings at each order of $\alpha'$ should be   independent of the background.  The background independence in the Double Field Theory formalism has been discussed in \cite{Hohm:2018zer}.

 The independent couplings in the effective action at a given order of $\alpha'$, are all gauge invariant couplings modulo  the field redefinitions, the total derivative terms and the Bianchi identities. For the closed spacetime manifolds, the field redefinitions at a given order of $\alpha'$ involve the most general gauge invariant terms at that order  \cite{Metsaev:1987zx}. The numbers of independent couplings  involving the metric, dilaton and the $B$-field at orders $\alpha',\alpha'^2,\alpha'^3$ are $8,60,872$, respectively \cite{Metsaev:1987zx,Garousi:2019cdn,Garousi:2020mqn}. The background independence assumption indicates that the coefficients of these couplings  are independent of the background. If one can fix them for a particular background, then they are valid for any other  background as well. On the other hand,  it has been proved  in \cite{Sen:1991zi,Hohm:2014sxa} that the dimensional reduction of the classical effective actions of the bosonic and heterotic string theories on a torus $T^d$ are  invariant under  global $O(d,d)$ transformations.   Hence, if one considers a particular background which includes one  circle,  compactifies the effective action on this circle and  ignores the Kaluza-Klein massive modes (dimensional reduction), then the lower dimensional action which includes all parameters of the original action,  must have the $O(1,1)$ symmetry. This symmetry may fix the couplings in the original action. Imposing this symmetry on the effective action of the bosonic string theory at orders $\alpha',\alpha'^2$, the coefficients of all independent couplings  have been found in \cite{Garousi:2019wgz,Garousi:2019mca} up to an overall factor. Imposing this symmetry on the  NS-NS couplings of the type II superstring theory at order $\alpha'^3$, the  coefficients of all independent couplings  have been found in \cite{Razaghian:2018svg,Garousi:2020gio,Garousi:2020lof} up to an overall factor. The resulting couplings must be valid for any other background. For example, the same couplings must be valid for the background which includes the compact sub-manifold $T^d$. The lower-dimensional action in this case must have the symmetry $O(d,d)$ \cite{Sen:1991zi,Hohm:2014sxa}. In fact, it has been shown in \cite{Garousi:2021ikb,Garousi:2021ocs} that the resulting couplings have exactly such symmetry. For the background which has the compact sub-manifold $T^2$, the lower dimensional action must have the symmetry $O(2,2)$. It has been shown in \cite{Eloy:2020dko} that the couplings at order $\alpha'$ have such symmetry.

For the open spacetime manifolds, it has been speculated  in \cite{Garousi:2021yyd} that the field redefinitions at a given order of $\alpha'$ involve only the restricted gauge invariant terms  which respect the boundary conditions in the least action principle. In the presence of boundary, the boundary conditions which are consistent with the above $O(1,1)$ or $O(d,d)$ symmetries, require the massless fields and their derivatives at order $m$ to be known on the boundary for  the effective actions at order $\alpha'^m$ \cite{Garousi:2021cfc}. The minimum numbers of independent couplings at order $\alpha'$  for the bosonic string theory and for the heterotic string theory  after truncating the Yang-Mills gauge fields, are 17  bulk couplings and 38  boundary couplings \cite{Garousi:2021yyd,Garousi:2021cfc}.  The background independence assumption in this case indicates that the coefficients of these parameters are independent of the background and are independent of the character of the boundary, \ie the coefficients in the bulk and  boundary couplings must be the same for both timelike and spacelike boundaries. Using the background independence assumption, these couplings in a particular minimal scheme  have been recently found for the timelike boundary to be \cite{Garousi:2021yyd}
\beqa
\bS_1
&=&-\frac{48a_1}{\kappa^2} \int_M d^{D}x \sqrt{-G} e^{-2\Phi}\Bigg[R^2_{\rm GB}+\frac{1}{24} H_{\alpha }{}^{\delta \epsilon } H^{\alpha \beta
\gamma } H_{\beta \delta }{}^{\varepsilon } H_{\gamma \epsilon
\varepsilon }-\frac{1}{8}  H_{\alpha \beta }{}^{\delta }
H^{\alpha \beta \gamma } H_{\gamma }{}^{\epsilon \varepsilon }
H_{\delta \epsilon \varepsilon }\nn\\&&\qquad\qquad +
R^{\alpha \beta }H_{\alpha }{}^{\gamma \delta } H_{\beta \gamma \delta }  -\frac{1}{12} R H_{\alpha
\beta \gamma } H^{\alpha \beta \gamma }  -\frac{1}{2} H_{\alpha }{}^{\delta \epsilon } H^{
\alpha \beta \gamma } R_{\beta \gamma \delta \epsilon
}\nn\\&&\qquad\qquad +4R \nabla_{\alpha }\Phi
\nabla^{\alpha }\Phi -16
 R^{\alpha \beta }\nabla_{\alpha }\Phi \nabla_{\beta
}\Phi \Bigg]\labell{ffinal}\\
\prt\!\!\bS^t_1&=&-\frac{48a_1}{\kappa^2}\int_{\prt M} d^{D-1}\sigma\sqrt{- g}  e^{-2\Phi}\Bigg[Q^t_2+ \frac{4}{3} n^{\alpha }
n^{\beta } \nabla_{\gamma }\nabla^{\gamma }K_{\alpha \beta
}-\frac{1}{6} H_{\beta \gamma \delta } H^{\beta \gamma \delta
} K^{\alpha }{}_{\alpha } +  H_{\alpha }{}^{\gamma \delta }
H_{\beta \gamma \delta } K^{\alpha \beta } \nn\\&&
\qquad\qquad +  H_{\alpha }{}^{\delta \epsilon } H_{\beta
\delta \epsilon } K^{\gamma }{}_{\gamma } n^{\alpha }
n^{\beta }  - 2 H_{\beta
}{}^{\delta \epsilon } H_{\gamma \delta \epsilon } n^{\alpha }
n^{\beta } n^{\gamma } \nabla_{\alpha }\Phi + 8 K^{\beta
}{}_{\beta } \nabla_{\alpha }\Phi \nabla^{\alpha }\Phi\nn\\&&\qquad\qquad - 16
K^{\gamma }{}_{\gamma } n^{\alpha } n^{\beta }
\nabla_{\alpha }\Phi \nabla_{\beta }\Phi - 16 K_{\alpha
\beta } \nabla^{\alpha }\Phi \nabla^{\beta }\Phi + \frac{32}{3}
n^{\alpha } n^{\beta } n^{\gamma } \nabla_{\alpha }\Phi
\nabla_{\beta }\Phi \nabla_{\gamma }\Phi \Bigg]\nn
\eeqa
where $n^\mu$ is the unit vector orthogonal to the boundary, $R^2_{\rm GB}$ is the Gauss-Bonnet gravity couplings   and $Q^t_2$ is the Chern-Simons couplings that for timelike boundary, \ie  $n^\mu n_\mu=1$, is given as \cite{Myers:1987yn,Garousi:2021cfc}
\beqa
Q^t_2&=&4\Bigg[K^\mu{}_{\mu}R-2K^{\mu\nu}R_{\mu\nu}-2K_\alpha{}^\alpha n^\mu n^\nu R_{\mu\nu}+2K^{\mu\nu}n^\alpha n^\beta R_{\alpha\mu\beta\nu}\nn\\&&\qquad-\frac{1}{3}(6K^\alpha{}_{\alpha}K_{\mu\nu}K^{\mu\nu}-2K^\mu{}_{\mu}K^\nu{}_{\nu}K^\alpha{}_{\alpha}-4K_{\mu}{}^{\nu}K_{\nu\alpha}K^{\alpha\mu})\Bigg]\labell{Q2t}
\eeqa
In above equations, $K_{\mu\nu}$ is the extrinsic curvature which is defined as $ K_{\mu\nu}=P^{\alpha}_{\ \mu}P^{\beta}_{\ \nu}\nabla_{(\alpha}n_{\beta)} $ where $P^{\mu\nu}$ is the first fundamental form which projects the spacetime tensors tangent to  the boundary. For the timelike boundary, the first fundamental form is defined as
$P^{\mu\nu}=G^{\mu\nu}-n^\mu n^\nu$  for which the extrinsic curvature becomes  $K_{\mu\nu}=\nabla_\mu n_\nu-n_\mu n_\alpha\nabla^\alpha n_\nu$.  In above equations, the metric in the covariant derivatives and in the curvatures  is the bulk metric $G_{\mu\nu}$. If one chooses the  overall factor $a_1$ to be $a_1=1/96$ ($a_1=1/192$), then  the bulk action becomes the Meissner  action of the bosonic (heterotic) string theory found in \cite{Meissner:1996sa}, up to a restricted field redefinition \cite{Garousi:2021yyd}. For the superstring theory $a_1=0$.

 For the spacelike boundary in which $n^\mu n_\mu=-1$, the first fundamental form is defined  as $P^{\mu\nu}=G^{\mu\nu}+n^\mu n^\nu$  for which the extrinsic curvature becomes  $K_{\mu\nu}=\nabla_\mu n_\nu+n_\mu n_\alpha\nabla^\alpha n_\nu$.  We expect this sign difference in the second term in the definition of the first fundamental form  changes the sign of all  terms which involve three unit vector $n^\mu$ or extrinsic curvature.
Therefore, when the spacetime has spacelike boundary, the background independence assumption of the effective action predicts  that the bulk action to be the same as the bulk action \reef{ffinal} for the spacetime which has  timelike boundary, whereas  the character independence  predicts the following boundary couplings for the spacelike boundary:
\beqa
\prt\!\!\bS^s_1&=&-\frac{48a_1}{\kappa^2}\int_{\prt M} d^{D-1}\sigma\sqrt{ g}  e^{-2\Phi}\Bigg[Q^s_2- \frac{4}{3} n^{\alpha }
n^{\beta } \nabla_{\gamma }\nabla^{\gamma }K_{\alpha \beta
}-\frac{1}{6} H_{\beta \gamma \delta } H^{\beta \gamma \delta
} K^{\alpha }{}_{\alpha } +  H_{\alpha }{}^{\gamma \delta }
H_{\beta \gamma \delta } K^{\alpha \beta } \nn\\&&
\qquad\qquad -  H_{\alpha }{}^{\delta \epsilon } H_{\beta
\delta \epsilon } K^{\gamma }{}_{\gamma } n^{\alpha }
n^{\beta }  + 2 H_{\beta
}{}^{\delta \epsilon } H_{\gamma \delta \epsilon } n^{\alpha }
n^{\beta } n^{\gamma } \nabla_{\alpha }\Phi + 8 K^{\beta
}{}_{\beta } \nabla_{\alpha }\Phi \nabla^{\alpha }\Phi\nn\\&&\qquad\qquad + 16
K^{\gamma }{}_{\gamma } n^{\alpha } n^{\beta }
\nabla_{\alpha }\Phi \nabla_{\beta }\Phi - 16 K_{\alpha
\beta } \nabla^{\alpha }\Phi \nabla^{\beta }\Phi - \frac{32}{3}
n^{\alpha } n^{\beta } n^{\gamma } \nabla_{\alpha }\Phi
\nabla_{\beta }\Phi \nabla_{\gamma }\Phi \Bigg]\labell{bSs}
\eeqa
where the terms involving three unit vector $n^\mu$ or extrinsic curvature have different sign compare to the timelike boundary couplings in \reef{ffinal}. In above equation, $Q^s_2$ is the Chern-Simons couplings for the spacelike boundary
\beqa
Q^s_2&=&4\Bigg[K^\mu{}_{\mu}R-2K^{\mu\nu}R_{\mu\nu}+2K_\alpha{}^\alpha n^\mu n^\nu R_{\mu\nu}-2K^{\mu\nu}n^\alpha n^\beta R_{\alpha\mu\beta\nu}\nn\\&&\qquad+\frac{1}{3}(6K^\alpha{}_{\alpha}K_{\mu\nu}K^{\mu\nu}-2K^\mu{}_{\mu}K^\nu{}_{\nu}K^\alpha{}_{\alpha}-4K_{\mu}{}^{\nu}K_{\nu\alpha}K^{\alpha\mu})\Bigg]\labell{Q2s}
\eeqa
Note that the terms with three extrinsic curvatures have different sign compare to the Chern-Simons couplings of the timelike boundary \reef{Q2t}.
In this paper, using the background independence method, we are going to calculate the bulk and boundary couplings for the spacetime manifold which has spacelike boundary and show that the resulting couplings  are exactly the same as the above couplings  which are predicted by the background/character independence assumption.

The outline  of the paper is as follows:  In section 2, we write the 17 independent bulk couplings and  the 38 independent   boundary couplings at order $\alpha'$ which have been found in \cite{Garousi:2021yyd,Garousi:2021cfc}.  In section 3, using the background independence assumption, we consider the background which has a spacelike boundary and one circle, and use the dimensional reduction to find the corresponding couplings in the base space. We then  impose the $O(1,1)$ symmetry on the reduced actions to constrain the parameters in the actions.     In section 4, we consider the background which has a spacelike boundary and the torus $T^d$, and use the cosmological  reduction to find the one-dimensional bulk action and the zero-dimensional boundary action. We then  impose the $O(d,d)$ symmetry on the resulting actions  to further constrain the remaining parameters.    The above two constraints fix the bulk action to be the bulk action in \reef{ffinal}, and fix the boundary action up to two parameters.  By requiring  the gravity couplings on the  boundary action to be  consistent with the  Chern-Simons couplings, the two boundary parameters are also fixed. We find that the final boundary action is exactly the same as \reef{bSs}.

\section{Independent couplings  at order $\alpha'$}

The  effective actions of   string theory on an open manifold has both bulk and boundary actions. At the sphere-level, these actions have the following $\alpha'$-expansion:
\beqa
\bS_{\rm eff}&=&\sum^\infty_{m=0}\alpha'^m\bS_m=\bS_0+\alpha' \bS_1 +\alpha'^2 \bS_2+\alpha'^3 \bS_3+\cdots \labell{seff}\\
\prt\!\!\bS_{\rm eff}&=&\sum^\infty_{m=0}\alpha'^m\prt\!\!\bS_m=\prt\!\!\bS_0+\alpha' \prt\!\!\bS_1+\alpha'^2 \prt\!\!\bS_2+\alpha'^3 \prt\!\!\bS_3 +\cdots \nn
\eeqa
The leading order  actions in the universal sector which includes metric, dilaton and B-field, in the string frame for both timelike and spacelike boundaries are
\beqa
\bS_0+\prt\!\!\bS_0
=-\frac{2}{\kappa^2}\Bigg[ \int d^{D}x \sqrt{-G} e^{-2\Phi} \left(  R + 4\nabla_{\mu}\Phi \nabla^{\mu}\Phi-\frac{1}{12}H^2\right)+ 2\int d^{D-1}\sigma\sqrt{| g|}  e^{-2\Phi}K\Bigg]\labell{baction}
\eeqa
where $G$ is determinant of the bulk metric $G_{\mu\nu}$ and boundary is specified by the functions $x^\mu=x^\mu(\sigma^{\tilde{\mu}})$. In the boundary term, $g$ is determinant of the induced metric on the boundary
\beqa
g_{\tilde{\mu}\tilde{\nu}}&=& \frac{\prt x^\mu}{\prt \sigma^{\tilde{\mu}}}\frac{\prt x^\nu}{\prt \sigma^{\tilde{\nu}}}G_{\mu\nu}\labell{indg}
\eeqa
and $K$ is the trace of the extrinsic curvature. The normal vector to the boundary is $n^\mu$. It is  outward-pointing (inward-pointing) if the boundary is spacelike (timelike).  % We consider in this paper only  the timelike boundary for which $n^\mu n_\mu=1$. %The boundary is timelike for which $n^\mu n_\mu=1$. The sign of the second term is minus if the boundary is timelike.  
Using the Double Field Theory formalism, it has been shown in \cite{Berman:2011kg} that the leading order effective action \reef{baction} can be written in   $O(D,D)$-invariant  form in terms of the generalized metric and dilaton.

At order $\alpha'$ these actions in terms of their Lagrangians  are
\beqa
\bS_1=-\frac{2}{\kappa^2}\int_M d^{D} x\sqrt{-G} e^{-2\Phi}\mathcal{L}_1 ;\qquad\prt\!\!\bS_1=-\frac{2}{\kappa^2}\int_{\prt M} d^{D-1} \sigma\sqrt{|g|} e^{-2\Phi}\prt\mathcal{L}_1
\eeqa
In general there are 41 gauge invariant couplings in the  bulk  Lagrangian. Removing the total derivative terms from the bulk to the boundary by using the Stokes's theorem and using the Bianchi identities, one can reduce the 41 couplings to  20  couplings. The most general field redefinitions reduce these couplings to 8 couplings \cite{Metsaev:1987zx}. However, in the presence of boundary one is not allowed to use the most general field redefinitions because they ruin  the boundary conditions required in the least action principle for the effective actions of string theory \cite{Garousi:2021cfc}.  The allowed field redefinitions at order $\alpha'$ requires the metric does not change, and the dilaton and B-field change to include only the first derivative of the massless fields. This restricted field redefinition has only three parameters. Hence, there are only 17  independent couplings in the bulk. The couplings  in a particular minimal scheme are \cite{Garousi:2021yyd}
\beqa
\mathcal{L}_1&= &  a_1 H_{\alpha }{}^{\delta \epsilon } H^{\alpha \beta \gamma }
H_{\beta \delta }{}^{\varepsilon } H_{\gamma \epsilon
\varepsilon } + a_2 H_{\alpha \beta }{}^{\delta } H^{\alpha
\beta \gamma } H_{\gamma }{}^{\epsilon \varepsilon } H_{\delta
\epsilon \varepsilon } + a_3 H_{\alpha }{}^{\gamma \delta }
H_{\beta \gamma \delta } R^{\alpha \beta } + a_4
R_{\alpha \beta } R^{\alpha \beta } \labell{L1bulk}\\&& + a_5
H_{\alpha \beta \gamma } H^{\alpha \beta \gamma } R +
a_6 R^2 + a_7 R_{\alpha \beta \gamma \delta }
R^{\alpha \beta \gamma \delta } + a_8 H_{\alpha
}{}^{\delta \epsilon } H^{\alpha \beta \gamma }
R_{\beta \gamma \delta \epsilon } + a_9 R
\nabla_{\alpha }\Phi \nabla^{\alpha }\Phi\nn\\&& + a_{10}
R^{\alpha \beta } \nabla_{\beta }\nabla_{\alpha }\Phi
+ a_{11} R_{\alpha \beta } \nabla^{\alpha }\Phi \nabla^{
\beta }\Phi + a_{12} \nabla_{\alpha }\Phi \nabla^{\alpha }\Phi
\nabla_{\beta }\Phi \nabla^{\beta }\Phi + a_{13} \nabla^{\alpha
}\Phi \nabla_{\beta }\nabla_{\alpha }\Phi \nabla^{\beta
}\Phi\nn\\&& + a_{14} \nabla_{\beta }\nabla_{\alpha }\Phi
\nabla^{\beta }\nabla^{\alpha }\Phi + a_{15} \nabla_{\alpha }H^{
\alpha \beta \gamma } \nabla_{\delta }H_{\beta \gamma
}{}^{\delta } + a_{16} H_{\alpha }{}^{\beta \gamma }
\nabla^{\alpha }\Phi \nabla_{\delta }H_{\beta \gamma
}{}^{\delta } + a_{17} \nabla_{\delta }H_{\alpha \beta \gamma }
\nabla^{\delta }H^{\alpha \beta \gamma }\nn
\eeqa
 where $a_1,\cdots, a_{17}$ are  17  background independent parameters.

The boundary of the spacetime has a unit normal vector $n^{\mu}$, hence, the boundary Lagrangian  $\prt {\cal L}_1$  should include this vector and its derivatives as well as the  bulk tensors. Since the field redefinition freedom has been already used in the bulk action, one is not allowed to use any field redefinition in the boundary action. Removing the boundary total derivative terms from the most general gauge invariant boundary couplings, and using the Bianchi identities and the identities corresponding to the unit vector, it has been shown in \cite{Garousi:2021cfc} that there are 38 independent couplings in the boundary action. For both timelike and spacelike boundaries, the couplings in a particular scheme  are \cite{Garousi:2021cfc}
\beqa
\prt \cL_1&=&b_{1}^{} H_{\beta \gamma \delta } H^{\beta \gamma \delta }
K^{\alpha }{}_{\alpha } + b_{2}^{} H_{\alpha }{}^{\gamma \delta
} H_{\beta \gamma \delta } K^{\alpha \beta } + b_{3}^{}
K_{\alpha }{}^{\gamma } K^{\alpha \beta } K_{\beta \gamma }
+ b_{4}^{} K^{\alpha }{}_{\alpha } K_{\beta \gamma } K^{\beta
\gamma } \nn\\&&+ b_{5}^{} K^{\alpha }{}_{\alpha } K^{\beta
}{}_{\beta } K^{\gamma }{}_{\gamma } + b_{6}^{} H_{\alpha
}{}^{\delta \epsilon } H_{\beta \delta \epsilon } K^{\gamma
}{}_{\gamma } n^{\alpha } n^{\beta } + b_{7}^{} H_{\alpha
\gamma }{}^{\epsilon } H_{\beta \delta \epsilon } K^{\gamma
\delta } n^{\alpha } n^{\beta } + b_{8}^{} K^{\alpha \beta }
R_{\alpha \beta } \nn\\&&+ b_{9}^{} K^{\gamma }{}_{\gamma }
n^{\alpha } n^{\beta } R_{\alpha \beta } + b_{10}^{}
K^{\alpha }{}_{\alpha } R + b_{11}^{} K^{\gamma \delta
} n^{\alpha } n^{\beta } R_{\alpha \gamma \beta
\delta } + b_{12}^{} H^{\beta \gamma \delta } n^{\alpha }
 \nabla_{\alpha }H_{\beta \gamma \delta }\nn\\&& + b_{13}^{} K^{\beta
\gamma } n^{\alpha } \nabla_{\alpha }K_{\beta \gamma } +
b_{14}^{} K^{\beta }{}_{\beta } n^{\alpha } \nabla_{\alpha
}K^{\gamma }{}_{\gamma } + b_{15}^{} n^{\alpha }
 \nabla_{\alpha }R + b_{16}^{} H_{\beta \gamma \delta }
H^{\beta \gamma \delta } n^{\alpha } \nabla_{\alpha }\Phi \nn\\&&+
b_{17}^{} K_{\beta \gamma } K^{\beta \gamma } n^{\alpha }
 \nabla_{\alpha }\Phi + b_{18}^{} K^{\beta }{}_{\beta }
K^{\gamma }{}_{\gamma } n^{\alpha } \nabla_{\alpha }\Phi +
b_{19}^{} H_{\beta }{}^{\delta \epsilon } H_{\gamma \delta
\epsilon } n^{\alpha } n^{\beta } n^{\gamma } \nabla_{\alpha
}\Phi \nn\\&&+ b_{20}^{} n^{\alpha } n^{\beta } n^{\gamma }
R_{\beta \gamma } \nabla_{\alpha }\Phi + b_{21}^{} n^{
\alpha } R \nabla_{\alpha }\Phi + b_{22}^{} K^{\beta
}{}_{\beta } \nabla_{\alpha }\Phi \nabla^{\alpha }\Phi +
b_{23}^{} n^{\alpha } n^{\beta } \nabla_{\alpha }\Phi
 \nabla_{\beta }K^{\gamma }{}_{\gamma }\nn\\&& + b_{24}^{} K^{\gamma
}{}_{\gamma } n^{\alpha } n^{\beta } \nabla_{\alpha }\Phi
 \nabla_{\beta }\Phi + b_{25}^{} n^{\alpha } n^{\beta }
 \nabla_{\beta } \nabla_{\alpha }K^{\gamma }{}_{\gamma } +
b_{26}^{} K^{\alpha \beta } \nabla_{\beta } \nabla_{\alpha
}\Phi\nn\\&& + b_{27}^{} K^{\gamma }{}_{\gamma } n^{\alpha }
n^{\beta } \nabla_{\beta } \nabla_{\alpha }\Phi + b_{28}^{} H_{
\alpha }{}^{\gamma \delta } H_{\beta \gamma \delta }
n^{\alpha } \nabla^{\beta }\Phi + b_{29}^{} n^{\alpha }
R_{\alpha \beta } \nabla^{\beta }\Phi + b_{30}^{}
K_{\alpha \beta } \nabla^{\alpha }\Phi \nabla^{\beta }\Phi \nn\\&&+
b_{31}^{} n^{\alpha } \nabla_{\alpha }\Phi \nabla_{\beta
}\Phi \nabla^{\beta }\Phi + b_{32}^{} n^{\alpha }
 \nabla_{\beta } \nabla_{\alpha }\Phi \nabla^{\beta }\Phi +
b_{33}^{} H_{\alpha }{}^{\delta \epsilon } n^{\alpha }
n^{\beta } n^{\gamma } \nabla_{\gamma }H_{\beta \delta
\epsilon } \nn\\&&+ b_{34}^{} n^{\alpha } n^{\beta } n^{\gamma }
 \nabla_{\alpha }\Phi \nabla_{\beta }\Phi \nabla_{\gamma
}\Phi + b_{35}^{} n^{\alpha } n^{\beta } n^{\gamma } \nabla_{
\alpha }\Phi \nabla_{\gamma } \nabla_{\beta }\Phi + b_{36}^{}
n^{\alpha } n^{\beta } n^{\gamma } \nabla_{\gamma
} \nabla_{\beta } \nabla_{\alpha }\Phi \nn\\&&+ b_{37}^{} n^{\alpha }
n^{\beta } \nabla_{\beta }K_{\alpha \gamma } \nabla^{\gamma
}\Phi + b_{38}^{} n^{\alpha } n^{\beta } n^{\gamma }
n^{\delta } \nabla_{\delta } \nabla_{\gamma }K_{\alpha \beta
}\labell{L1boundary}
\eeqa
 where $b_1,\cdots, b_{38}$ are 38 background independent parameters. These parameters, however, depend on the character of boundary.  They have been found in \cite{Garousi:2021yyd} for the timelike boundary. In the following sections we consider the boundary to be spacelike.

\section{Background  with  sub-manifold $S^{(1)}$}

We have used the gauge symmetries corresponding to the massless fields to write the independent couplings in the bulk  action \reef{L1bulk} and in the boundary action \reef{L1boundary}. The parameters in these actions are independent of the backgrounds. In general, there is no global symmetry in the universal sector of  string theory at the critical dimension to be used for fixing these parameters. However, for some specific backgrounds which have compact sub-manifolds, the compactified  actions may have some global symmetries after ignoring the Kaluza-Klein massive modes (dimensional reduction).
Since the parameters in the actions \reef{L1bulk} and \reef{L1boundary} appears also in the lower dimensional actions, one can use the symmetry of the lower dimensional actions to fix these parameters. In this section we consider the  background with sub-manifold $S^{(1)}$.  That is,  we choose the open manifold to have  the structure $M^{(D)}=M^{(D-1)}\times S^{(1)}$, $\prt M^{(D)}=\prt M^{(D-1)}\times S^{(1)}$. The manifold $M^{(D)}$ has coordinates $x^\mu=(x^a,y)$ and  its boundary  $\prt M^{(D)}$ has coordinates $\sigma^{\tilde{\mu}}=(\sigma^\ta, y)$ where $y$ is the coordinate of the circle $S^{(1)}$.  The boundary in the base space  is specified by the functions $x^a=x^a(\sigma^{\ta})$. The dimensionally reduced  action then should have the $O(1,1)$ symmetry. To simplify the calculation, we consider the $Z_2$-subgroup of the $O(1,1)$-group.

 The reduction of the effective actions on the circle  $S^{(1)}$ should then be invariant under the $Z_2$-transformations, up to some total derivative terms on the boundary \cite{Garousi:2019xlf}, \ie
 \beqa
 S_{\rm eff}(\psi)+\prt S_{\rm eff}(\psi)&=&S_{\rm eff}(\psi')+\prt S_{\rm eff}(\psi')\labell{TT}
 \eeqa
where  $S_{\rm eff}$ and  $\prt S_{\rm eff}$  are  the reductions of the bulk action $\!\!\bS_{eff}$ and boundary action $\prt\!\! \bS_{\rm eff}$, respectively. In above equation $\psi$ represents all the  massless fields in the base space which are defined in the following Kaluza-Klein reductions \cite{Maharana:1992my}:
 \beqa
&&G_{\mu\nu}=\left(\matrix{\bg_{ab}+e^{\varphi}g_{a }g_{b }& e^{\varphi}g_{a }&\cr e^{\varphi}g_{b }&e^{\varphi}&}\right),B_{\mu\nu}= \left(\matrix{\bb_{ab}+\frac{1}{2}b_{a }g_{b }- \frac{1}{2}b_{b }g_{a }&b_{a }\cr - b_{b }&0&}\right),\nn\\&& \Phi=\bar{\phi}+\varphi/4\,,\quad n^{\mu}=(n^a,0)\labell{reduc}
\eeqa
and $\psi'$ represents its transformation under the $Z_2$-transformations  or the T-duality transformations. At the leading order of $\alpha'$, the T-duality transformations  are the Buscher rules \cite{Buscher:1987sk,Buscher:1987qj}. To order $\alpha'$, they are the Buscher rules  and the corrections at order $\alpha'$ which do not ruin the boundary  conditions of the least action principle in the base space. They are \cite{Garousi:2021yyd}
\beqa
&&\varphi'= -\varphi+\alpha'\Delta\vp
\,\,\,,\,\,g'_{a }= b_{a }+\alpha'e^{\vp/2}\Delta g_a\,\,\,,\,\, b'_{a }= g_{a }+\alpha'e^{-\vp/2}\Delta b_a \,\,\,,\,\,\nn\\
&&\bg_{ab}'=\bg_{ab} \,\,\,,\,\,\bH_{abc}'=\bH_{abc}+\alpha'\Delta\bH_{abc} \,\,\,,\,\,  \bar{\phi}'= \bar{\phi}+\alpha'\Delta\bphi\,\,\,,\,\, n_a'=n_a\labell{T22}
\eeqa
where the corrections  $\Delta \vp, \Delta b_a,\Delta g_a,\Delta \bphi$ contain all contractions of the  massless fields in the base space at order $\alpha'$ which involve only the first derivative of the massless fields.  The correction $\Delta \bH_{abc}$ is related to the corrections  $\Delta g_a$,  $\Delta b_a$ through the following relation:
\beqa
\Delta\bH_{abc}&=&\tilde H_{abc}-3e^{-\vp/2}W_{[ab}\Delta b_{c]}-3e^{\vp/2}\Delta g_{[a}V_{bc]}
\eeqa
where $\tilde H_{abc}$ is a $U(1)\times U(1)$ gauge invariant closed 3-form at order $\alpha'$ which is odd under parity.   It has the following terms:
\beqa
\tilde H_{abc}&=&e_1\prt_{[a}W_{b}{}^dV_{c]d}+e_2\prt_{[a}\bH_{bc]d}\nabla^d\vp\labell{tH}
\eeqa
where $e_1,e_2$ and the coefficients in the corrections   $\Delta \vp, \Delta b_a,\Delta g_a,\Delta \bphi$ are  parameters that the $Z_2$-symmetry of the effective action  should fix them. The above transformations should also  form the $Z_2$-group  \cite{Garousi:2019wgz}. In the above equation, $V_{ab}$ is field strength of the $U(1)$ gauge field $g_{a}$, \ie $V_{ab}=\prt_{a}g_{b}-\prt_{b}g_{a}$, and $W_{\mu\nu}$ is field strength of the $U(1)$ gauge field $b_{a}$, \ie $W_{ab}=\prt_{a}b_{\nu}-\prt_{b}b_{a}$. The    three-form $\bH$ is defined as $\bH_{abc}=\hat{H}_{abc}-\frac{3}{2}g_{[a}W_{bc]}-\frac{3}{2}b_{[a}V_{bc]}$ where the three-form  $\hat{H}$ is field strength of the two-form $\bb_{ab}  $ in \reef{reduc}.

In \cite{Garousi:2021cfc}, it has been shown that  the constraint \reef{TT} can be written as two separate constraints. One for the bulk couplings and the other one for the boundary couplings. These constraints for the couplings at order $\alpha'$ are \cite{Garousi:2021cfc}
\beqa
 S_1(\psi)-S_1(\psi'_0)-\Delta S_0-\frac{2}{\kappa^2}\int d^{D-1}x\sqrt{-\bg}\nabla_a (A_1^a e^{-2\bphi})&=&0\nn\\
\prt S_1(\psi)-\prt S_1(\psi'_0)-\Delta\prt S_0+T_1(\psi)+\frac{2}{\kappa^2}\int d^{D-2}\sigma\sqrt{\tg}\,n_{a} A_1^a e^{-2\bphi}&=&0\labell{S11b}
\eeqa
where $\bg$ is the determinant of the base space metric $\bg_{ab}$ and $\tg$ is the determinant  of the induced base space metric on its boundary, \ie
\beqa
\tg_{\ta\tb}&=&\frac{\prt x^{a}}{\prt \sigma^{\ta}}\frac{\prt x^{b}}{\prt \sigma^{\tb}}\bg_{ab}\labell{gtatb}
 \eeqa
In equation \reef{S11b}, $\psi_0'$ is the transformation of the base space field $\psi$ under the Buscher rules,   $A_1^a$ is a  vector made of the  massless fields in the base space at order $\alpha'$ with arbitrary coefficients, and   $T_1(\psi)$ is the most general total derivative terms in the boundary at order $\alpha'$, \ie
\beqa
T_1(\psi)=-\frac{2}{\kappa^2}\int_{\prt M^{(D-1)}}d^{D-2}\sigma \sqrt{|\tg |} n_{a}\nabla_{b}(e^{-2\bphi}F_1^{ab})
\labell{bstokes1}
\eeqa
where  $ F_1^{ab} $ is an  antisymmetric  tensor constructed  from the  massless fields in the base space at order $\alpha'$ with arbitrary coefficients.
In the equation \reef{S11b}, $\Delta S_0$, $\Delta\prt S_0$ are the Taylor expansions  of the reduction of the leading order actions \reef{baction} at order $\alpha'$,
\beqa
S_0(\psi'_0+\alpha'\psi'_1)&=&S_0(\psi'_0)+\alpha'\Delta S_0+\cdots\nn\\
\prt S_0(\psi'_0+\alpha'\psi'_1)&=&\prt S_0(\psi'_0)+\alpha'\Delta \prt S_0+\cdots\labell{DS0}
\eeqa
where dots represent some terms at higher orders of $\alpha'$ in which we are not interested in this paper.

The first constraint in \reef{S11b} involves only the bulk fields that their reductions are given in \cite{Garousi:2019mca}. The second constraint involves the bulk fields and the boundary  extrinsic curvature. The reduction of the extrinsic curvature and its first and second derivatives for the timelike boundary are calculated in \cite{Akou:2020mxx}. We have checked explicitly that they are valid for spacelike boundary as well. Using these reductions and
following the same steps as those in \cite{Garousi:2021cfc}, one finds that the $Z_2$-symmetry fixes the bulk Lagrangian \reef{L1bulk} to be  the same as the one has been found in \cite{Garousi:2021yyd}, \ie
\beqa
\mathcal{L}_1&= &a_{1}^{} H_{\alpha }{}^{\delta \epsilon } H^{\alpha \beta
\gamma } H_{\beta \delta }{}^{\varepsilon } H_{\gamma \epsilon
\varepsilon } + (3 a_{1}^{} + \frac{1}{64} a_{10}^{} +
\frac{1}{64} a_{11}^{}) H_{\alpha \beta }{}^{\delta }
H^{\alpha \beta \gamma } H_{\gamma }{}^{\epsilon \varepsilon }
H_{\delta \epsilon \varepsilon } -  \frac{1}{16} a_{11}^{}
H_{\alpha }{}^{\gamma \delta } H_{\beta \gamma \delta }
R^{\alpha \beta }\nn\\&& + (\frac{1}{4} a_{10}^{} +
\frac{1}{4} a_{11}^{}) R_{\alpha \beta }
R^{\alpha \beta } + \frac{1}{192} a_{11}^{} H_{\alpha
\beta \gamma } H^{\alpha \beta \gamma } R -
\frac{1}{16} a_{11}^{} R^2 + 24 a_{1}^{}
R_{\alpha \beta \gamma \delta } R^{\alpha
\beta \gamma \delta }\nn\\&& + (-36 a_{1}^{} -  \frac{1}{8} a_{10}^{}
-  \frac{1}{16} a_{11}^{}) H_{\alpha }{}^{\delta \epsilon } H^{
\alpha \beta \gamma } R_{\beta \gamma \delta \epsilon
} -  \frac{1}{4} a_{11}^{} R \nabla_{\alpha }\Phi
\nabla^{\alpha }\Phi + a_{10}^{} R^{\alpha \beta }
\nabla_{\beta }\nabla_{\alpha }\Phi\nn\\&& + a_{11}^{}
R_{\alpha \beta } \nabla^{\alpha }\Phi \nabla^{\beta
}\Phi + a_{10}^{} \nabla_{\beta }\nabla_{\alpha }\Phi
\nabla^{\beta }\nabla^{\alpha }\Phi -  \frac{1}{16} a_{10}^{}
\nabla_{\alpha }H^{\alpha \beta \gamma } \nabla_{\delta
}H_{\beta \gamma }{}^{\delta }\nn\\&& + \frac{1}{8} a_{10}^{}
H_{\alpha }{}^{\beta \gamma } \nabla^{\alpha }\Phi
\nabla_{\delta }H_{\beta \gamma }{}^{\delta } + (8 a_{1}^{} +
\frac{1}{24} a_{10}^{} + \frac{1}{48} a_{11}^{})
\nabla_{\delta }H_{\alpha \beta \gamma } \nabla^{\delta
}H^{\alpha \beta \gamma }\labell{fL1}
\eeqa
 and the boundary Lagrangian \reef{L1boundary} for  the spacelike  boundary to be
 \beqa
\prt \cL_1&=&b_{1}^{} H_{\beta \gamma \delta } H^{\beta \gamma \delta }
K^{\alpha }{}_{\alpha } + \frac{1}{16} (-2 a_{10}^{} -
a_{11}^{}) H_{\alpha }{}^{\gamma \delta } H_{\beta \gamma
\delta } K^{\alpha \beta } + b_{11}^{} K_{\alpha }{}^{\gamma }
K^{\alpha \beta } K_{\beta \gamma } \nn\\&&+ \frac{1}{4} \bigl(-
a_{11}^{} - 2 (48 b_{1}^{} + b_{17}^{})\bigr) K^{\alpha
}{}_{\alpha } K_{\beta \gamma } K^{\beta \gamma } +
\frac{1}{12} (a_{11}^{} + 96 b_{1}^{} - 2 b_{18}^{}) K^{\alpha
}{}_{\alpha } K^{\beta }{}_{\beta } K^{\gamma }{}_{\gamma }\nn\\&& -
\frac{1}{2} b_{19}^{} H_{\alpha }{}^{\delta \epsilon } H_{\beta
\delta \epsilon } K^{\gamma }{}_{\gamma } n^{\alpha }
n^{\beta } + (a_{10}^{} + \frac{1}{2} a_{11}^{} + 12 b_{12}^{})
K^{\alpha \beta } R_{\alpha \beta } \nn\\&&+ \frac{1}{2}
(a_{10}^{} - 48 b_{1}^{} + 24 b_{12}^{} -  b_{17}^{} + 4 b_{19}^{})
K^{\gamma }{}_{\gamma } n^{\alpha } n^{\beta }
R_{\alpha \beta } + (- \frac{1}{8} a_{11}^{} - 12
b_{1}^{}) K^{\alpha }{}_{\alpha } R\nn\\&& + b_{11}^{}
K^{\gamma \delta } n^{\alpha } n^{\beta } R_{\alpha
\gamma \beta \delta } + b_{12}^{} H^{\beta \gamma \delta } n^{
\alpha } \nabla_{\alpha }H_{\beta \gamma \delta } +
(\frac{1}{48} a_{11}^{} - 2 b_{1}^{}) H_{\beta \gamma \delta }
H^{\beta \gamma \delta } n^{\alpha } \nabla_{\alpha }\Phi \nn\\&&+
b_{17}^{} K_{\beta \gamma } K^{\beta \gamma } n^{\alpha }
\nabla_{\alpha }\Phi + b_{18}^{} K^{\beta }{}_{\beta }
K^{\gamma }{}_{\gamma } n^{\alpha } \nabla_{\alpha }\Phi +
b_{19}^{} H_{\beta }{}^{\delta \epsilon } H_{\gamma \delta
\epsilon } n^{\alpha } n^{\beta } n^{\gamma } \nabla_{\alpha
}\Phi \nn\\&&+ (- a_{10}^{} -  \frac{1}{2} a_{11}^{} - 24 b_{12}^{} +
b_{17}^{} - 4 b_{19}^{}) n^{\alpha } n^{\beta } n^{\gamma }
R_{\beta \gamma } \nabla_{\alpha }\Phi + (-
\frac{1}{4} a_{11}^{} + 24 b_{1}^{}) n^{\alpha } R
\nabla_{\alpha }\Phi\nn\\&& - 48 b_{1}^{} K^{\beta }{}_{\beta }
\nabla_{\alpha }\Phi \nabla^{\alpha }\Phi - 2 (48 b_{1}^{} +
b_{18}^{}) K^{\gamma }{}_{\gamma } n^{\alpha } n^{\beta }
\nabla_{\alpha }\Phi \nabla_{\beta }\Phi\nn\\&& + \frac{1}{2}
\bigl(4 a_{10}^{} + a_{11}^{} + 48 (-2 b_{1}^{} +
b_{12}^{})\bigr) K^{\alpha \beta } \nabla_{\beta
}\nabla_{\alpha }\Phi \nn\\&&+ (a_{10}^{} + \frac{1}{2} a_{11}^{} +
24 b_{12}^{} -  b_{17}^{} + 4 b_{19}^{}) K^{\gamma }{}_{\gamma }
n^{\alpha } n^{\beta } \nabla_{\beta }\nabla_{\alpha }\Phi +
\frac{1}{8} a_{10}^{} H_{\alpha }{}^{\gamma \delta } H_{\beta
\gamma \delta } n^{\alpha } \nabla^{\beta }\Phi \nn\\&&+
\frac{1}{2} (a_{11}^{} - 96 b_{1}^{}) n^{\alpha }
R_{\alpha \beta } \nabla^{\beta }\Phi + a_{11}^{} K_{
\alpha \beta } \nabla^{\alpha }\Phi \nabla^{\beta }\Phi + (-
a_{11}^{} + 96 b_{1}^{}) n^{\alpha } \nabla_{\alpha }\Phi
\nabla_{\beta }\Phi \nabla^{\beta }\Phi\nn\\&& + (a_{11}^{} - 96
b_{1}^{}) n^{\alpha } \nabla_{\beta }\nabla_{\alpha }\Phi
\nabla^{\beta }\Phi + \frac{1}{8} (2 a_{10}^{} + a_{11}^{} - 2
b_{11}^{} + 48 b_{12}^{}) H_{\alpha }{}^{\delta \epsilon }
n^{\alpha } n^{\beta } n^{\gamma } \nabla_{\gamma }H_{\beta
\delta \epsilon } \nn\\&&-  \frac{2}{3} \bigl(a_{11}^{} - 2 (96
b_{1}^{} + b_{18}^{})\bigr) n^{\alpha } n^{\beta } n^{\gamma }
\nabla_{\alpha }\Phi \nabla_{\beta }\Phi \nabla_{\gamma
}\Phi \labell{L12}\\&&- 2 (a_{10}^{} + 48 b_{1}^{} + 24 b_{12}^{} -  b_{17}^{} + 4
b_{19}^{}) n^{\alpha } n^{\beta } n^{\gamma } \nabla_{\alpha
}\Phi \nabla_{\gamma }\nabla_{\beta }\Phi + b_{38}^{}
n^{\alpha } n^{\beta } n^{\gamma } n^{\delta }
\nabla_{\delta }\nabla_{\gamma }K_{\alpha \beta }\nn
 \eeqa
which is not the same as its corresponding timelike Lagrangian found in \cite{Garousi:2021yyd}. The sign of some of the parameters are changed compare to the timelike case. We have  imposed the identities corresponding to the unit vector in the base space, by writing it as
\beqa
n^a&=&-\frac{\prt_a f}{\sqrt{|\prt_b f\prt^b f|}}
\eeqa
where $f$ is the function that specifies the spacelike boundary, \ie $\prt_af\prt^a f=-|\prt_a f\prt^a f|$.

The bulk Lagrangian has three parameters $a_1, a_{10},a_{11}$ and the boundary Lagrangian has two bulk parameters $a_{10},a_{11}$ and 7 boundary parameters $b_1,b_{11},b_{12},b_{17},b_{18},b_{19},b_{38}$. Since not all parameters are fixed up to an overall factor, in the next section we consider another  background.

\section{Background  with  sub-manifold $T^{(d)}$}

 In this section, we consider the  background which has the  sub-manifold $T^{(d)}$.  That is,  the open manifold has the structure $M^{(D)}=M^{(1)}\times T^{(d)}$, $\prt M^{(D)}=\prt M^{(1)}\times T^{(d)}$. The base space manifold $M^{(1)}$ has time coordinate $t$, hence,  its boundary is spacelike boundary.  The compactification on this background has massless modes as well as  infinite tower of massive Kaluza-Klein modes.  If one ignores the massive Kaluza-Klein modes (cosmological reduction), and uses the appropriate one-dimensional field redefinitions, then the cosmological  action  should have the $O(d,d)$ symmetry,
 \ie
 \beqa
 S^c_{\rm eff}(\psi)+\prt S^c_{\rm eff}(\psi)&=&S^O_{\rm eff}(\psi')+\prt S^O_{\rm eff}(\psi')\labell{TTc}
 \eeqa
where  $S^c_{\rm eff}$ and  $\prt S^c_{\rm eff}$  are  the cosmological reductions of the bulk action $\!\!\bS_{eff}$ and boundary action $\prt\!\! \bS_{\rm eff}$, respectively. In above equation $\psi$ represents all the  massless fields in the base space, \ie
 \beqa
G_{\mu\nu}=\left(\matrix{- n^2(t)& 0&\cr 0&G_{ij}(t)&}\right),\, B_{\mu\nu}= \left(\matrix{0&0\cr0&B_{ij}(t)&}\right),\,  2\Phi=\phi+\frac{1}{2}\log\det(G_{ij})\labell{creduce}\eeqa
The lapse function $n(t)$ can also be fixed to $n=1$. This function at the boundary is the unit vector orthogonal to the boundary.
On the right-hand side of equation \reef{TTc}, $\psi'$ represents their appropriate higher-derivative field redefinitions. The effective actions on the right-hand side must be  invariant under $O(d,d)$-transformations.

In the absence of boundary, it has been shown in \cite{Hohm:2015doa,Hohm:2019jgu} that there are field redefinitions, including the lapse function, in which the non-local cosmological action which involves higher time-derivatives become local action which involves only the first time-derivative of the generalized metric.  In the presence of the boundary, one should not use the field redefinitions for the lapse function because this function at the boundary  represents the unit normal vector on the boundary. Moreover, in the presence of the boundary, the field redefinitions should be restricted to those which do not ruin the boundary conditions in the least action principle in the  base space \cite{Garousi:2021yyd}. In the presence of boundary,  there might be the field redefinitions that left intact  the laps function  and do not ruin the boundary conditions, however,  the  action in that scheme may involve the first derivative of  the generalized metric as well as the first derivative of the one-dimensional dilaton, \ie the action may still become local.  On the other hand, for the local action, one expects the usual boundary condition in the least action principle in which only the values of the massless fields are known on the boundary. Hence, the boundary action should not include the derivative of the massless fields, \ie as it has been speculated in \cite{Garousi:2021cfc}, the boundary action must be zero in that particular scheme. Hence, in that scheme, the  constraint \reef{TTc}  becomes
 \beqa
 S^c_{\rm eff}(\psi)+\prt S^c_{\rm eff}(\psi)&=&S^O_{\rm eff}(\psi')\labell{TTc1}
 \eeqa
 It has been shown in \cite{Garousi:2021yyd} that there is such scheme  at order $\alpha'$.

The above constraint at each order of $\alpha'$ produces two constraints. One bulk and one boundary constraints. At the leading order of $\alpha'$, they are
\beqa
S_0^c(\psi)-\frac{2}{\kappa^2}\int dt \frac{d}{dt}(\cI_0e^{-\phi})&=&S_0^O(\psi)\nn\\
\prt S_0^c(\psi)+\frac{2}{\kappa^2}\cI_0e^{-\phi}&=&0\labell{twoc0}
\eeqa
The second terms in the first equation is a total derivative term at the two derivative order. For a particular $\cI_0$, the bulk constraint produces the following $O(d,d)$-invariant action  \cite{Veneziano:1991ek,Meissner:1991zj,Hohm:2015doa}:
\beqa
S_0^O&=&-\frac{2}{\kappa^2 }\int dt e^{-\phi}\Bigg[-\dP^2-\frac{1}{8}\tr(\dS^2)\Bigg]\labell{S0}
\eeqa
where $\cS$ is the generalized metric. Taking into the account the appropriate $\cI_0$ from the bulk constraint, one finds the boundary constraint \reef{twoc0} satisfies automatically \cite{Garousi:2021cfc}.

The  constraint \reef{TTc1} at  order  $\alpha'$ produces the following two constraints:
\beqa
S_1^c(\psi)-\Delta S_0^O(\psi)-\frac{2}{\kappa^2}\int dt \frac{d}{dt}(\cI_1e^{-\phi})&=&S_1^O(\psi)\nn\\
\prt S_1^c(\psi)+\frac{2}{\kappa^2}\cI_1e^{-\phi}&=&0\labell{twoc}
\eeqa
where $\Delta S_0^O(\psi)$ is the Taylor expansion of the leading order cosmological action \reef{S0} at order $\alpha'$, \ie
\beqa
S_0^O(\psi+\alpha'\psi_1')&=&S_0^O(\psi)+\alpha'\Delta S_0^O(\psi)+\cdots
\eeqa
It has been shown in \cite{Garousi:2021yyd} that the bulk constraint in \reef{twoc} is satisfies  when there are  the following relations between the bulk parameters $a_1,a_{10},a_{11}$:
\beqa
a_{11}&\!\!=\!\!&-384 a_1,\,\,\,a_{10}\,=\,0\labell{a11011}
\eeqa
The corresponding $O(d,d)$-invariant  action is the cosmological action that has been  found in \cite{Meissner:1996sa}, \ie
 \beqa
S_1^O(\psi)&=&-\frac{2}{\kappa^2 }24a_1\int dt e^{-\phi}\Bigg[\frac{1}{16}\tr(\dS^4)-\frac{1}{64}(\tr(\dS^2))^2+\frac{1}{2}\tr(\dS^2)\dP^2-\frac{1}{3}\dP^4\Bigg]\labell{action2}
 \eeqa
The corresponding  total derivative terms are   the following:
\beqa
\cI_1&=&24 a_{1}^{}  \dB_{i}{}^{k} \dB^{ij}
\dG_{jk} +12a_1\dG^{i}{}_{i} \dG_{jk} \dG^{jk}
-6a_1 \dB_{ij} \dB^{ij} \dG^{k}{}_{k}\nn\\&& - 6a_1 \dG^{i}{}_{i} \dG^{j}{}_{j} \dG^{k}{}_{k} -24a_1  \dB_{ij} \dB^{ij} \dP + 24a_1  \dG^{i}{}_{i} \dP^2 + 32a_1 \dP^3  \labell{I1}
\eeqa
The corresponding  field redefinitions that involve only the first derivative of the massless fields have been also found  in \cite{Garousi:2021yyd}. However, since they do not appear in the boundary constraint in \reef{twoc},  we are not interested in them. Inserting the relations \reef{a11011} into the bulk Lagrangian \reef{fL1}, one reproduces the Lagrangian \reef{ffinal}, as expected.

The one-dimensional  reduction of the timelike boundary couplings \reef{L12} is the following:
\beqa
\prt S_1^c&=&-\frac{2 }{\kappa^2 }e^{-\phi}\Big[\frac{1}{4} \bigl( -96 a_1 -  (b_{11}^{} -
12 b_{12}^{})\bigr) \dB_{i}{}^{k} \dB^{ij} \dG_{jk} + \frac{1}{4}
\bigl(-192 a_1 -  (b_{11}^{} - 12
b_{12}^{})\bigr) \dG_{i}{}^{k} \dG^{ij} \dG_{jk}\nn\\&& +(36 a_1+6b_1-3b_{12}) \dG^{i}{}_{i} \dG_{jk} \dG^{jk} - 6 a_1 \dB_{ij} \dB^{ij} \dG^{k}{}_{k}-6 a_1 \dG^{i}{}_{i} \dG^{j}{}_{j} \dG^{k}{}_{k}\labell{ds1c}\\&& -
\frac{1}{2} (24 a_1 + 6 b_{1}^{} +
b_{19}^{}) \dB_{ij} \dB^{ij} \dP + (60 a_1 + 9 b_{1}^{} -6b_{12}- \frac{1}{2} b_{19}^{}) \dG_{ij} \dG^{ij} \dP \nn\\&&
+24 a_1 \dG^{i}{}_{i} \dP^2 + \frac{1}{6}
(96 a_1 - 24 b_{1}^{} -  b_{18}^{}) \dP^3 + \frac{1}{4} ( 192 a_1 +  b_{11}^{} - 12 b_{12}^{}) \dB^{ij}
\ddB_{ij}\nn\\&& + \frac{1}{4} ( 192 a_1 + b_{11}^{} -
12 b_{12}^{}) \dG^{ij} \ddG_{ij} + \frac{1}{2} (-192 a_1+ 24 b_{12}^{} -  b_{17}^{} + 4 b_{19}^{}) \dP
\ddP \Big]\nn
\eeqa
where we have also used the relations \reef{a11011}. The above action is not invariant under the $O(d,d)$ transformations.
If one includes  $\cI_1$ which is given in \reef{I1}, one can choose the boundary parameters  such that the result becomes invariant. For the following relations between the parameters:
\beqa
b_{11}=-96a_1+24b_1,\,\,\,\,b_{12}=8a_1%-\frac{1}{24}a_{10}
+2b_1,&&b_{19}=-12b_1+\frac{1}{4}b_{17}\labell{b1219}
\eeqa
The boundary action becomes  $O(d,d)$-invariant  which involves the first derivative of the dilaton, \ie
\beqa
\prt S_1^c(\psi)+\frac{2}{\kappa^2}\cI_1e^{-\phi}&=&-\frac{2 }{\kappa^2 }e^{-\phi}\Big[(12 a_1 + 3 b_{1}^{} -\frac{1}{8}b_{17})( \dB_{ij} \dB^{ij} +\dG_{ij} \dG^{ij})\dP   \nn\\&&
\qquad\qquad\quad-
(16 a_1 +4 b_{1}^{} +\frac{1}{6}  b_{18}^{}) \dP^3  \Big]\nn
\eeqa
The boundary constraint in \reef{twoc} then dictates the following relations:
\beqa
b_{17}=96a_1+24b_1,&& b_{18}=-96a_1%+\frac{1}{2}a_{10}
-24b_1\labell{bb}
\eeqa
  The above relations \reef{b1219} and \reef{bb}, then reduce the 7 boundary parameters in \reef{L12} to 2 parameters $b_1,b_{38}$. Note that the coupling with coefficient $b_{38}$ is invariant under the $O(1,1)$ and $O(d,d)$ transformations.

For the spacetime manifolds which have boundary,  both the bulk and boundary actions should satisfy the least action principle, \ie $\delta(\!\!\bS_1+\prt\!\!\bS_1)=0$ with the appropriate boundary condition on the massless fields. Since the bulk action has at most the  term with two derivatives, the variation of the bulk action satisfies   $\delta \!\!\bS_1=0$ using the assumption that the values of the massless fields and their first derivatives are known on the boundary \cite{Garousi:2021cfc}. The variation of the boundary action produces variation of the second derivatives of the massless fields which are not zero on the boundary for the effective action at order $\alpha'$. However, the  parameters $b_1,b_{38}$,   can not be fixed because the non-zero variations are total derivative terms on the boundary which are zero.
In fact inserting the relations \reef{a11011}, \reef{b1219} and \reef{bb} into the boundary action \reef{L12}, one finds the variation of the resulting boundary action  against  the metric variation produces the following terms:
\beqa
&& -24(4a_1 +b_{1}^{}) \prt^{\alpha }\Phi \prt^{\beta }f
P^{\gamma \delta } \nabla_{\alpha }\nabla_{\beta
}\delta G_{\gamma \delta }
 -24(4a_1+b_1) \prt^{\alpha }\Phi
\prt_{\alpha }f \prt^{\beta }f \prt^{\gamma }f P^{\delta \epsilon }
\nabla_{\beta }\nabla_{\gamma }\delta G_{\delta \epsilon }\labell{varg}
\eeqa
where  we have used the assumption that the variation of metric and its first derivative, and their tangent derivatives   are zero, \ie $\delta G_{\alpha\beta}=\prt_\mu\delta G_{\alpha\beta}=0$ and $P^{\mu\nu}\prt_\mu\prt_\gamma\delta G_{\alpha\beta}=0$.
On the other hand, if one considers the following antisymmetric tensor:
\beqa
\cF_1^{\alpha\beta}&=&(96a_1+24b_1)n^\mu n^{[\alpha}\prt^{\beta]}\Phi(\nabla_\mu\delta G^\nu{}_\nu-\nabla_\nu\delta G_\mu{}^\nu)
\eeqa
Then its corresponding  boundary total derivative term, \ie
\beqa
\int_{\prt M^{(D)}}d^{D-1}\sigma \sqrt{g } \,n_{\alpha}\nabla_{\beta}(e^{-2\Phi}\mathcal{F}_1^{\alpha\beta})\labell{totb}
\eeqa
would cancel the variations \reef{varg}. Similar cancellations happen for the variations of the boundary action against the dilaton and B-field.

We fix the remaining boundary parameters $b_1,\,b_{38}$  by noting that the boundary couplings include the structures as those in the Chern-Simons form. Hence, we fix the remaining parameters in the boundary action  such that  the gravity couplings in the boundary  include the Chern-Simons form. The Chern-Simons form has the following gravity couplings for the spacelike  boundary\cite{Myers:1987yn}:
\beqa
Q^s_2&=&4\Bigg[K^\mu{}_{\mu}\tR-2K^{\mu\nu}\tR_{\mu\nu}-\frac{1}{3}(3K^\alpha{}_{\alpha}K_{\mu\nu}K^{\mu\nu}-K^\mu{}_{\mu}K^\nu{}_{\nu}K^\alpha{}_{\alpha}-2K_{\mu}{}^{\nu}K_{\nu\alpha}K^{\alpha\mu})\Bigg]\labell{Q20}
\eeqa
where $\tR_{\mu\nu}$ and $\tR$ are curvatures that are constructed from the induced metric \reef{indg}.  Using the  following Gauss-Codazzi  relations for the spacelike boundary:
 \beqa
\tR_{\alpha\beta}&=&P_{\alpha\mu}P_{\beta\nu}R^{\mu\nu}+n^\mu n^\nu R_{\alpha\mu\beta\nu}+K_{\alpha\mu}K_{\beta}{}^{\mu}-K_{\alpha\beta}K_\mu {}^{\mu}\nn\\
\tR&=&R+2n^\mu n^\nu R_{\mu\nu}+K_{\mu\nu}K^{\mu\nu}-K_\mu {}^\mu K_\nu {}^\nu
\eeqa
and the identity $n^\mu K_{\mu\nu}=0$, one can rewrite $Q^s_2$ in terms of the spacetime curvatures, \ie \reef{Q2s}.
For the spacelike boundary, there is also the following  identity:
\beqa
n^{\alpha } n^{\beta } n^{\gamma } n^{\delta }
\nabla_{\delta }\nabla_{\gamma }K_{\alpha \beta }&=&2 K_{\alpha }{}^{\gamma } K^{\alpha \beta } K_{\beta
\gamma }-n^{\alpha } n^{\beta }
\nabla_{\gamma }\nabla^{\gamma }K_{\alpha \beta }\labell{iden}
\eeqa
which can be verified by writing both sides in the local frame and in terms of the function $f$, \ie
\beqa
n^\mu&=&-\frac{\prt_\mu f}{\sqrt{|\prt_\nu f\prt^\nu f|}}
\eeqa
The identity \reef{iden}
 is the same as the corresponding identity in the timelike boundary \cite{Garousi:2021cfc} in which the terms which have three $n^\alpha$ or extrinsic curvature, have different sign. However, the term which has five $n^\alpha$ or extrinsic curvature, has the same sign.

Using the identity \reef{iden},  one finds the gravity couplings in the boundary action become the same as the couplings in $Q^s_2$ for the following relations:
\beqa
b_1=-4a_1,\,b_{38}=32 a_1\labell{b1b11}
\eeqa
In fact, inserting the relations \reef{a11011}, \reef{b1219}, \reef{bb} and \reef{b1b11}  into the boundary action \reef{L12}, one finds the boundary couplings \reef{bSs} dictated by the background/character independence of the effective actions at the critical dimension.

The boundary action for the non-null boundaries can then be written as
\beqa
\prt\!\!\bS_1&\!\!\!\!=\!\!\!\!\!&-\frac{48a_1}{\kappa^2}\int d^{D-1}\sigma\sqrt{|g|}  e^{-2\Phi}\Bigg[Q_2+ \frac{4}{3}n^2 n^{\alpha }
n^{\beta } \nabla_{\gamma }\nabla^{\gamma }K_{\alpha \beta
}-\frac{1}{6} H_{\beta \gamma \delta } H^{\beta \gamma \delta
} K^{\alpha }{}_{\alpha } +  H_{\alpha }{}^{\gamma \delta }
H_{\beta \gamma \delta } K^{\alpha \beta } \nn\\&&
\qquad\quad +  n^2H_{\alpha }{}^{\delta \epsilon } H_{\beta
\delta \epsilon } K^{\gamma }{}_{\gamma } n^{\alpha }
n^{\beta }  - 2 n^2H_{\beta
}{}^{\delta \epsilon } H_{\gamma \delta \epsilon } n^{\alpha }
n^{\beta } n^{\gamma } \nabla_{\alpha }\Phi + 8 K^{\beta
}{}_{\beta } \nabla_{\alpha }\Phi \nabla^{\alpha }\Phi\labell{fff}\\&&\qquad\quad - 16
n^2K^{\gamma }{}_{\gamma } n^{\alpha } n^{\beta }
\nabla_{\alpha }\Phi \nabla_{\beta }\Phi - 16 K_{\alpha
\beta } \nabla^{\alpha }\Phi \nabla^{\beta }\Phi + \frac{32}{3}
n^2n^{\alpha } n^{\beta } n^{\gamma } \nabla_{\alpha }\Phi
\nabla_{\beta }\Phi \nabla_{\gamma }\Phi \Bigg]\nn
\eeqa
where $n^2=n^\mu n_\mu$, and the Chern-Simons density is
\beqa
Q_2&=&4\Bigg[K^\mu{}_{\mu}R-2K^{\mu\nu}R_{\mu\nu}-2n^2K_\alpha{}^\alpha n^\mu n^\nu R_{\mu\nu}+2n^2K^{\mu\nu}n^\alpha n^\beta R_{\alpha\mu\beta\nu}\nn\\&&\qquad-\frac{1}{3}n^2(6K^\alpha{}_{\alpha}K_{\mu\nu}K^{\mu\nu}-2K^\mu{}_{\mu}K^\nu{}_{\nu}K^\alpha{}_{\alpha}-4K_{\mu}{}^{\nu}K_{\nu\alpha}K^{\alpha\mu})\Bigg]
\eeqa
The boundary action  has terms which have one and five $n^\alpha$ and/or $K_{\alpha\beta}$. At the higher orders of $\alpha'$, one expects the boundary action to have terms with $1,5,9,13,\cdots$ unit vector $n^\alpha$ and/or $K_{\alpha\beta}$. Hence, to find the boundary actions at higher orders of $\alpha'$, one may first find the couplings for the timelike boundary, and then inserts $n^2=1$ in the couplings which have $3,7,11,\cdots$ unit vector $n^\alpha$ and/or $K_{\alpha\beta}$, to produce terms with $1,5,9,13,\cdots$ unit vector $n^\alpha$ and/or $K_{\alpha\beta}$. The result then would be valid for the spacelike boundary as well. It would be interesting to find the boundary couplings at order $\alpha'^2$ to check this speculation. 

The boundary action for  the null boundary may  be obtained by
treating the null boundary as a limit of a sequence of non-null boundaries \cite{Parattu:2015gga}. Using this method,  the coupling on the null boundary at the leading order has been found in \cite{Parattu:2015gga} by taking the appropriate limit of the non-null boundary coupling in \reef{baction}. One may  use  this method to find the boundary couplings at order $\alpha'$ for the  null boundary by taking the limit of the non-null boundary couplings \reef{fff}.

Having found the boundary couplings \reef{fff} corresponding to the bulk action \reef{ffinal} at order $\alpha'$, one may try to write them in manifest $O(D,D)$-invariant form, as has been done in \cite{Berman:2011kg} for the leading order effective action.  Even for the closed spacetime manifold,  it is hard to write the effective action at order $\alpha'$ in terms of the generalised metric because the conventional 2D-dimensional Riemann  curvature does not transform covariantly under the generalized diffeomorphisms \cite{Hohm:2010xe,Hohm:2011si,Jeon:2011cn,Coimbra:2011nw}. However, this action has been written in $O(D,D)$-invariant form using the generalized frame \cite{Marques:2015vua,Baron:2017dvb}. It would be interesting to write the bulk and boundary actions \reef{ffinal}, \reef{fff} in the  duality manifest actions  in terms of the generalized frame.

% \vskip .3 cm
%{\bf Acknowledgements}:   This work is supported by Ferdowsi University of Mashhad.% under grant  1/49736(1398/02/17).

%\newpage

\end{document}